# BENCHMARKING EXERCISES FOR GRANULAR FLOWS

Antoine Lucas
*Institut de Physique du Globe de Paris, Université Paris Diderot, France*

Anne Mangeney
*Institut de Physique du Globe de Paris, Université Paris Diderot, France
and Institute for Nonlinear Sciences, UCSD, San Diego, USA*

François Bouchut
*Département de Mathématiques et Applications, École Normale Supérieure, France*

Marie-Odile Bristeau
*Institut National de Recherche en Informatique et Automatique, Le Chesnay, France*

Daniel Mège
*Laboratoire de Planétologie et de Géodynamique, Université de Nantes, France*

**Abstract:** *For the 2007 International Forum on Landslide Disaster Management framework, our team performed several numerical simulations on both theoretical and natural cases of granular flows. The objective was to figure out the ability and the limits of our numerical model in terms of reproduction and prediction. Our benchmarking exercises show that for almost all the cases, the model we use is able to reproduce observations at the field scale. Calibrated friction angles are almost similar to that used in other models and the shape of the final deposits is in good agreement with observation. However, as it is tricky to compare the dynamics of natural cases, these exercises do not allow us to highlight the good ability to reproduce the behavior of natural landslides. Nevertheless, by comparing with analytical solution, we show that our model presents very low numerical dissipation due to the discretization and to the numerical scheme used. Finally, in terms of mitigation and prediction, the different friction angles used for each cases figure out the limits of using such model as long as constitutive equations for granular media are not known.*

## INTRODUCTION

**Model Description**
Simulation is performed using the numerical model, hereafter called *Shaltop-2d*, resulting from a long term collaboration between Département de Mathématiques et Applications (DMA), École Normale Supérieure (ENS, Paris) and Institut de Physique du Globe de Paris (IPGP) in France (Bouchut et al. 2003; Bouchut 2004; Bouchut & Westdickenberg 2004; Mangeney et al. 2005, 2007). This model was developed after former studies performed by our group in the context of a collaboration between IPGP, ENS-DMA and INRIA. A first model was developed based on the model developed for shallow river flows by Audusse et al. (2000) and Bristeau et al. (2001).

Mangeney et al. (2003) extended this model based on a kinetic scheme to the classical Savage-Hutter model for granular flows over sloping topography with a Coulomb-type basal friction involving either a constant friction coefficient or the empirical flow rule proposed by Pouliquen (1999). Finally, the projection of the gravity field on the reference frame linked to the topography was performed during the PhD of M. Pirulli together with the active/passive earth pressure coefficient (Pirulli et al. 2007) and several basal friction laws have been tested on real events (Pirulli 2004; Pirulli & Mangeney 2007). This model after called RASH$^{3D}$ by Pirulli et al. (2007) is used in the Benchmarking Exercice by Pirulli and Scavia. The main advantage of this model is the unstructured finite element mesh best suited to deal with complex topography. However, the numerical method is only of the first order and the friction, the projection of the gravity field and the introduction of the earth pressure coefficients are not compatible with the preservation of steady state and do not respect all the properties required by the numerical resolution of the equations (see Bouchut 2004 for details about numerical methods to solve these problems). Furthermore, this kinetic model (i.e. RASH$^{3D}$ model) is based on the Savage-Hutter equations developed in a reference frame linked to the topography including only the curvature terms in the $x$ and $y$ direction. It is noted that the aforementioned problems are shared with most of the numerical models proposed in the literature.

Contrary to the classical approaches, the new model *Shaltop-2d* is based on the equations developed by Bouchut et al. (2003) and Bouchut & Westdickenberg (2004) developed in a fixed cartesian reference frame with the thin layer approximation (TLA) imposed in the direction perpendicular to the topography (see Figure 1). The rigourous asymptotic analysis makes it possible for the first time to account for the whole curvature tensor. Furthermore, the numerical method is of the second order requiring less refined grid to reach the same precision. The numerical method is based on the work of Bouchut (2004) and preserve the steady states as well as other requirements related to the resolution of hyperbolic equations.

The overall idea is to develop the equations in a fixed reference frame ($x, y, z$), for example horizontal/vertical, as opposed to the equations developed by Hutter and his co-workers in a variable reference frame linked to the topography. However, the shallowness assumptions are still imposed in the local reference frame ($X, Y, Z$) linked to the topography (Figure 1). Indeed, to satisfy the hydrostatic assumption for shallow flow over inclined topography, it is the acceleration normal to the topography that must be neglected compared to the gradient of the pressure normal to the topography. The reference frame is shown in Figure 1. The 2D horizontal coordinate vector is $x = (x, y) \in R^2$ and the topography is described by the scalar function $b(x, y)$ with a 3D unit upward normal vector:

$$\vec{n} \equiv (-s, c) \in \Re^2 \times \Re \qquad [1]$$

where,

$$s = \frac{\nabla_{xy} z}{\sqrt{1 + \|\nabla_{xy} z\|^2}} \qquad [2]$$

and

$$c = \frac{1}{\sqrt{1 + \|\nabla_{xy} z\|^2}} \qquad [3]$$

where $\nabla_{xy}$ is the gradient of the topography in both directions and $\theta$ is the angle between $\vec{n}$ and the vertical.

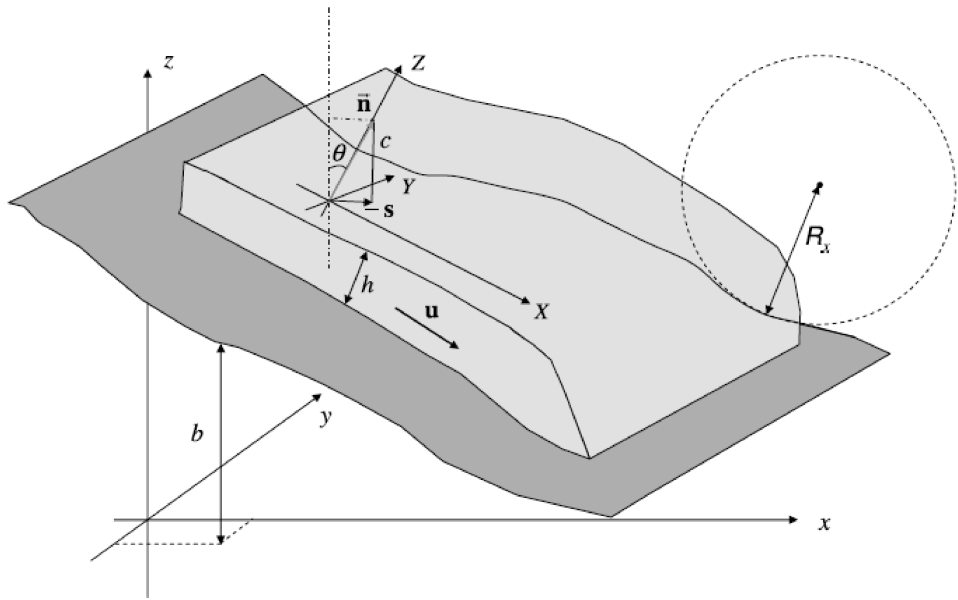

Figure 1: Frames and variables used in the model (after Mangeney et al. 2007)

In the horizontal/vertical Cartesian coordinate and for a inclined plane, formulation of the equations reduce to:

$$\partial_t h + c \cdot \partial_x (hu) + \partial_y (hv) = 0 \quad [4]$$

$$\partial_t (u) + cu \cdot \partial_x (u) + v \cdot \partial_y (u) + c \cdot \partial_x (ghc) = -g \sin\theta + \tilde{f}_{fx} \quad [5]$$

$$\partial_t (v) + cu \cdot \partial_x (v) + v \cdot \partial_y (v) + \partial_y (ghc) = \tilde{f}_{fy} \quad [6]$$

where $h$ is the depth-thickness of the flow, $g$ is the vertical acceleration and $\mathbf{u}(u, v)$ is the depth-average velocity. The two-dimensional Saint-Venant system with topography and friction is written:

$$\begin{cases} \partial_t h + \partial_x (hu) + \partial_y (hu) = 0 \\ \partial_t (hu) + \partial_x (hu^2 + gh^2/2) + \partial_y (huv) + hg\partial_x b = hf_x \\ \partial_t (hv) + \partial_x (huv) + \partial_y (hv^2 + gh^2/2) + hg\partial_y b = hf_y \end{cases} \quad [7]$$

where $b(x, y)$ represents the 2D topography.

Debris avalanche is treated here as a single-phase, dry granular flow with Coulomb-type behavior. The transition between a static state and flowing state is modeled by introducing a threshold allowing or not the material to flow. The friction forces $f(x, y, t) = (f_x, f_y)$ (see Eqs. [5] and [6]) must satisfy

$$\| f(x,y,t) \| \leq g\mu \qquad [8]$$

$$u(x,y,t) \neq 0 \Rightarrow f(x,y,t) = -g\mu \frac{u(x,y,t)}{\| u(x,y,t) \|} \qquad [9]$$

The friction coefficient is either supposed to be constant (Coulomb friction law) or depending on the thickness $h$ and the Froude number of the flow $u/(gh)^{1/2}$ (hereafter called Pouliquen flow rule):

(1) Coulomb friction law defined as follow:

$$\mu = \tan\delta \qquad [10]$$

where $\delta$ is the constant friction angle.

(2) Pouliquen friction emprirical rule (Pouliquen 1999):

$$\mu(\|u\|,h) = \tan\delta_1 + (\tan\delta_2 - \tan\delta_1)\exp\left(-\beta \frac{h}{d}\frac{\sqrt{gh}}{\|u\|}\right) \qquad [11]$$

where $\delta_1$ and $\delta_2$ are characteristic friction angles of the material, $d$ is a length scale of the order of a grain diameter, and $\beta = 0.136$ is a dimensionless parameter as it has been proposed by Pouliquen (1999). Basically, the friction parameter is increasing for small thickness h and high velocity u. The Coulomb type friction law applied on the averaged media has been tested by comparing depth-averaged model *Shaltop-2d* with discrete element simulations (Mangeney et al. 2006) on the collapse of granular column of variable aspect ratio ($a = H_i/R_i$, where $H_i$ and $R_i$ are the intial thickness and radius of the granular column). The main result was that vertical acceleration has to be included in the model if aspect ratio $a > 1$ are dealt with.

*Shaltop-2d* has been used successfully to simulate laboratory experiments of granular collapse (Mangeney-Castelnau et al. 2005), levees-channel formation (Mangeney et al. 2007) as well as real 3D avalanches in a natural context as it has been recently shown for large Martian landslides (Lucas & Mangeney 2007).

**Computing Ressources**
We benefit of the computing facilities from the Metacentre de Données et de Calcul Parallèle of IPGP:

*CPU Server*
- Server Bi-Xeon 5160 3.00GHz
  - Mem: 3Gb
  - OS: Linux Fedora 8
  - Compiler: Intel Fortran 10.1
- IBM® e325 server bi-CPU AMD Opteron 256
  - Mem: 4Gb
  - OS: Linux RedHat Entreprise 3
  - Compiler: pgf95 - The Portland Group Inc. Fortran 90/95 compiler

*Parallel Computing Cluster*
- 64 x IBM x 3550 bi-CPU Intel Xeon Quad-Core E5420 (512 cores)
  - Mem: 512Gb
  - OS: RedHat EL release 5.1
  - Compiler: Intel Fortran 10.1

Time of computation could rise up to several days on the IBM e325 depending on the grid space increment and the time of the simulation. Parallelization of the code (using MPI) has been carried out by P. Stoclet at IPGP.

**Softwares Used**
DEM (Digital Elevation Model) processing (e.g. gridding) were performed with Surfer (Golden software). Visualization and rendering have been carry out under GMT (Wessel & Smith 1991) and MatLab.

**Cases Studied**
As previously mentioned, Shaltop−2d is appropriate for dry granular monophasic flow such as landslides, rock avalanches and dry sand flows. In this context, we focused our runs on the following cases even though water could be present in the real landslides:

(1) Dam break scenario from analytical solution
(2) Laboratory test of dry sand flow from USGS
(3) Shum Wan landslide, Hong Kong
(4) Fei Tsui Road landslide, Hong Kong
(5) Frank slide, Canada

**DAM-BREAK SCENARIO**

**Description of the Scenario**
This test makes it possible to calibrate the numerical modeling with analytical solution. The scenario involves movement of mass in a wide inclined channel, triggered by sudden removal of dam (Figure 2). From the analytical solution (AS) derived by Mangeney et al. (2000), profiles of the debris flow at any given time $t$ are:

$$h(t) = \begin{cases} H & ; \quad x \leq -x_r \\ \frac{1}{9g\cos\theta}(2c_0 - \frac{x}{t} - \frac{mt}{2})^2 & ; -x_r < x < x_l \\ 0 & ; \quad x_l \leq x \end{cases} \quad [12]$$

with

$$m = g(\cos\theta \tan\delta - \sin\theta) \quad [13]$$

$$c_0 = \sqrt{gH\cos\theta} \quad [14]$$

$$x_l = 2c_0 t - \frac{1}{2}mt^2 \quad [15]$$

$$x_r = c_0 t + \frac{1}{2} m t^2 \qquad [16]$$

where $H$ is the initial height of debris mass, $g$ is the gravity, $\theta$ is the slope angle and $\delta$ is the angle of friction.

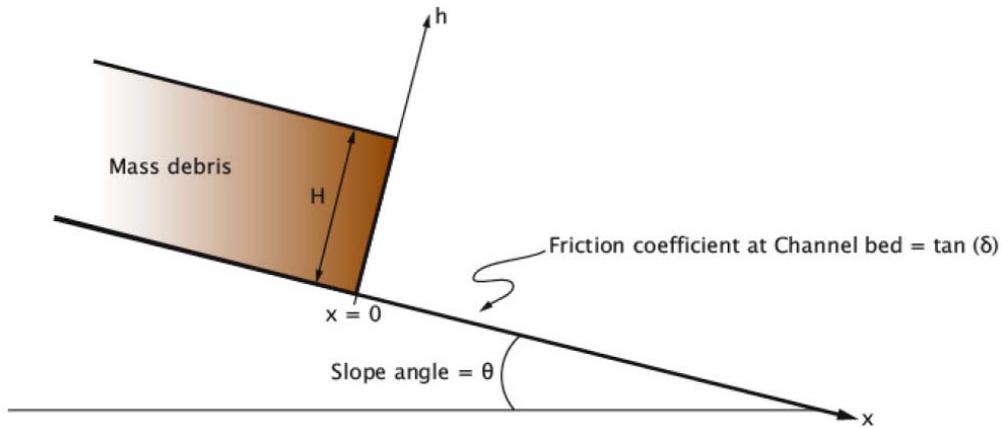

Figure 2: Initial conditions of dam-break scenario. $H$ is the initial height of the debris mass, $\theta$ is the slope angle and $\delta$ is the angle of friction at channel bed.

We perform test using the following settings (Figure 3):

- Initial height of the mass: 10m
- Slope angle: 30°
- Friction angle: 25°

**Results**
*Shaltop-2d* is able predict very well the analytical solution (Figure 3).

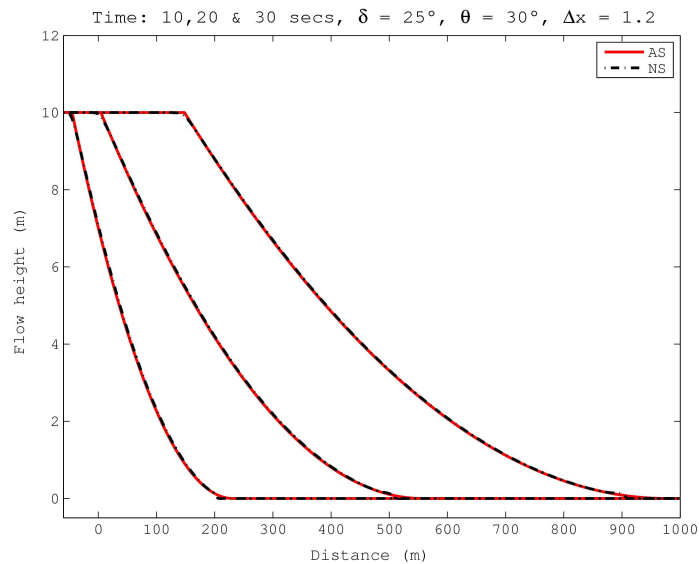

Figure 3: Evolution of the flow height for both analytical and numerical solutions respectively named AS and NS. Using a reasonable space increment Δx, *Shaltop-2d* is able to reproduce very well the analytical solution.

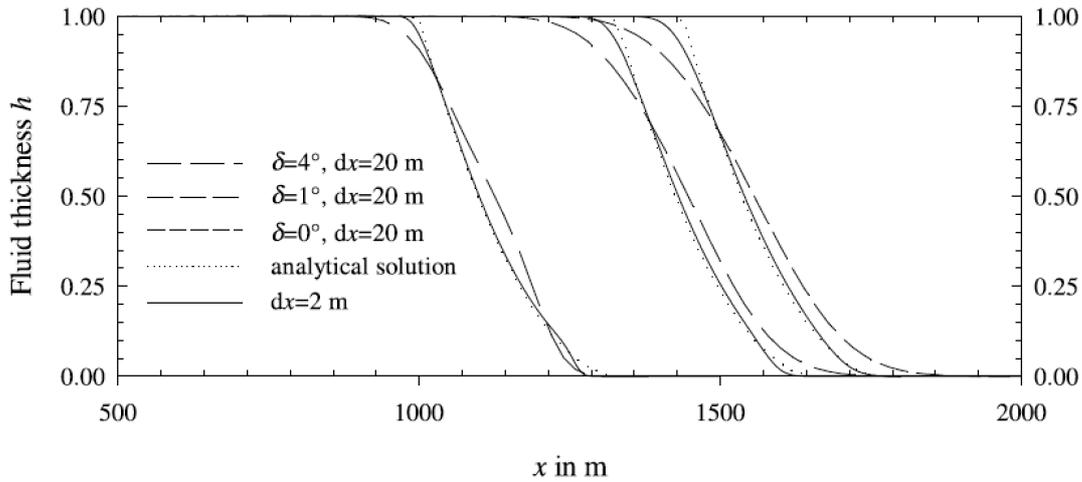

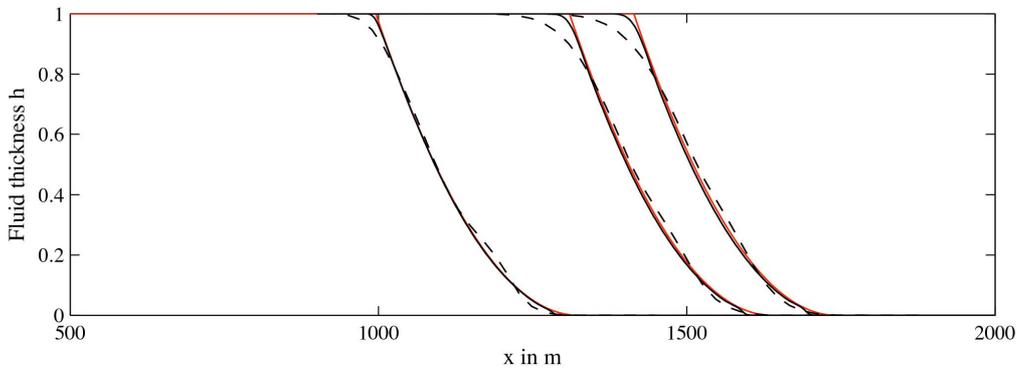

Figure 4: (a) Comparaison between RASH3D model with analytical solution (figure from Mangeney et al. 2003). (b) Using the same parameters, we compare here *Shaltop-2d* with anlytical solution.

In Figure 4, we show that, for the same parameters, *Shaltop-2d* fits better with the analytical solutions than RASH$^{3D}$ model (Mangeney et al. 2003). For $\delta = 4°$, $\theta = 5°$ and $\Delta x = 20$, *Shaltop-2d* repoduces better the runout distance predicted by the analytical solution (Figure 4). The overestimation of this runout distance in the kinetic model is related to numerical diffusion. As a result, the calibration of the kinetic model in order to fit the right runout distance would lead to friction angle almost one degree higher than the real friction angle. In the simulation of real avalanches over complex topography, numerical diffuusion is expected to generate larger runout distances (i.e. larger value of the fitted friction angle) using kinetic scheme or similar schemes of the first order than that obtained with *Shaltop-2d*. A series of numerical test on real topography with increasing discretization is therefore necessary if the models (i.e. the fitted friction coefficients) are to be compared on real events.

## DRY SAND BEHAVIOR WITHIN A 3D CHANNEL

### Description of the Simulation

Simulation of the experimental results of Iverson et al. (2004) dealing with dry granular flows over irregular 3D channel (see Figure 5) is performed here.
The basal topography has the following characteristics (Figure 5):

- Mean angle of sloping flume bed: 31.6°
- Headgate aperture width: 12 cm
- X location of headgate: 8.858 cm
- Location of urethane topographic insert: entire flume width (20cm), and covering from $x$ = 9.14 cm to $x$ = 39.12 cm
- The right triangular prism that spans the full width of the flume (20cm) is approximately $h_0$ = 4.35 cm thick (measured vertically) at $x$ = 8.858 cm (headgate) and tapers to 0 cm thick upflume from the headgate at x ≈ 1.9 cm.

We interpolate (using kriging algorithms) the DEM so as to get a space increment $\Delta x = \Delta y$ = 0.24 cm.

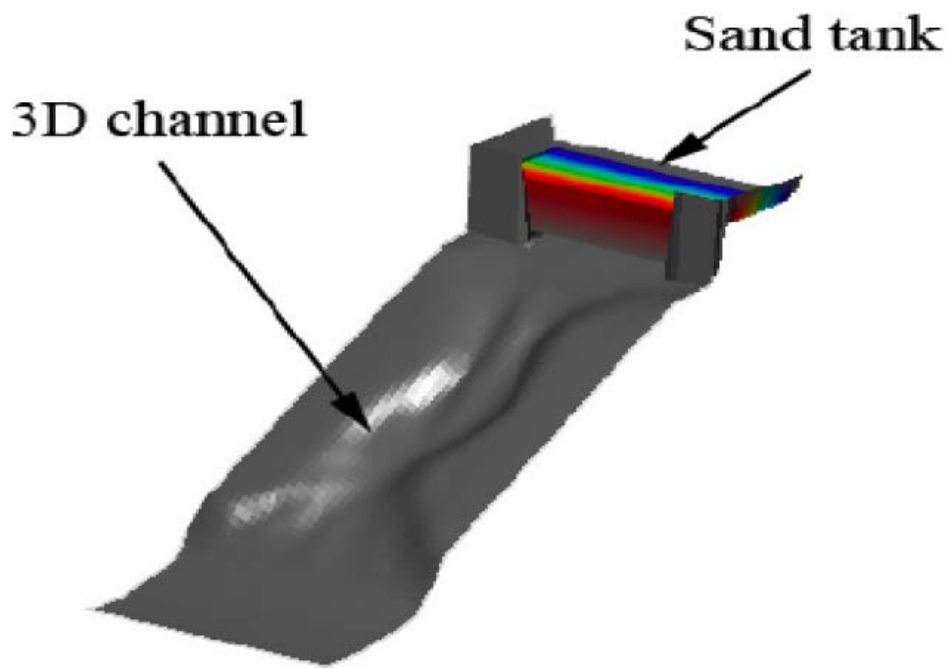

Figure 5: 3D irregular channel with the initial mass (in color)

Results obtained using a constant friction coefficient all over the topography are obviously not accurate because substrates with different roughness were used in the tank and in the irregular topography. Numerical simulation usign $\delta = 29°$ all over the numerical domain is however shown for comparison (see Figure 7). The bottom part of the mass spreading is quite well reproduced by the code, whereas the top part does not fit with Iverson et al. (2004) observations. Mass staying in the tank is overestimated within the sand tank at the end of the simulation. As consequences, 15% of the bulk mass don't participate in the spreading.

As mentioned in Iverson et al. (2004), two kind of materials are used in this 3D channel. The tank and the bottom parts are made of Formica with a bed friction angle $\delta_f$ = 23.5° and the channel part with urethane characterized by a bed friction angle $\delta_u$ = 19.8°. We thus adapt our code so as to take into account the variation of the bed friction angle due to the different floor (Formica and urethane). Results are obviously significantly improved as shown in Figure 8. Mass profile is more alike experimental observations (top of Figure 6).

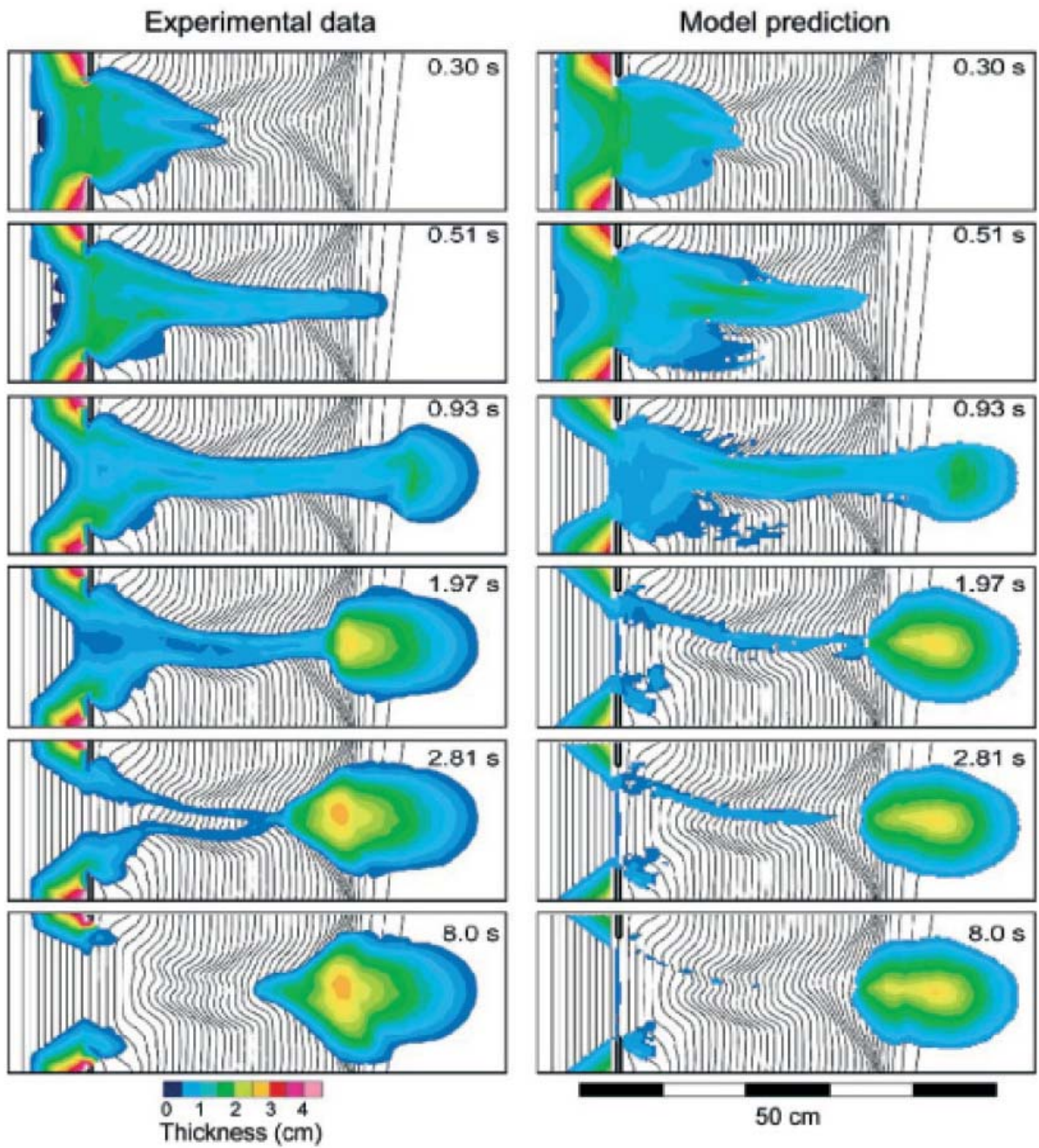

Figure 6: Evolution of the dry granular media across from Iverson et al. (2004)

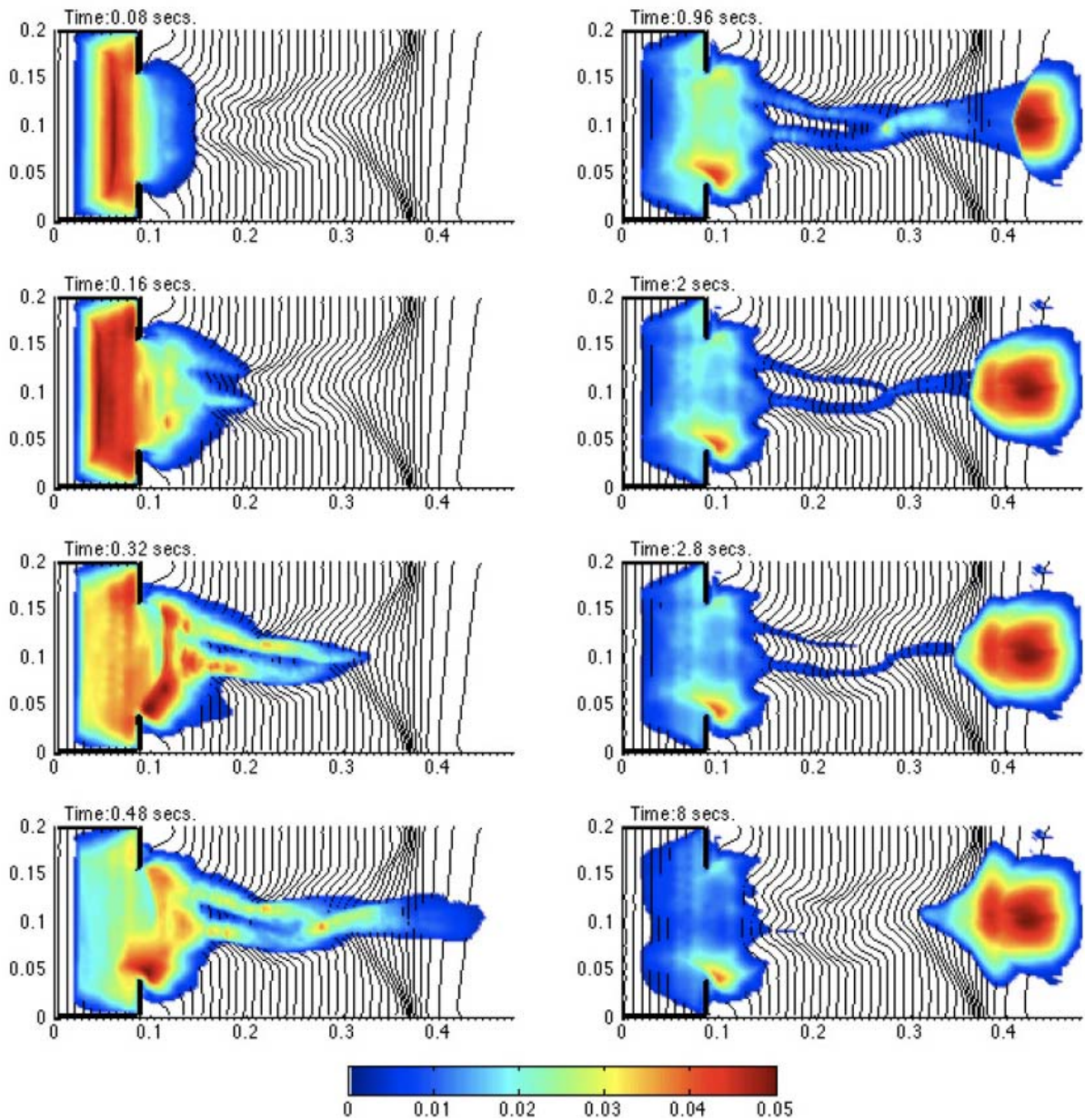

Figure 7: Evolution of the dry granular media accross an irregular 3D channel using a single friction angle δ = 29° all over the numerical domain (DEM courtesy Iverson et al. 2004)

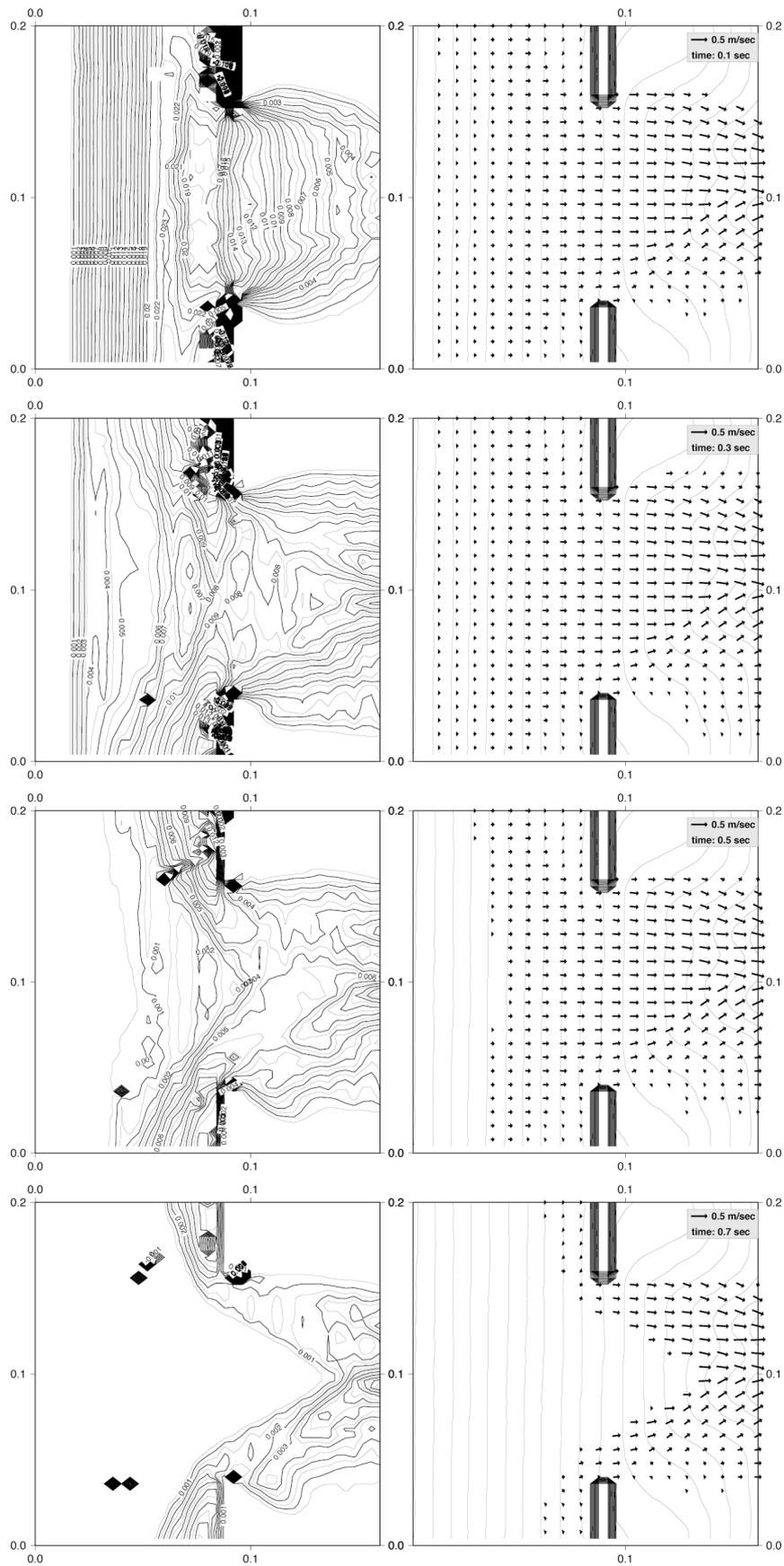

Figure 8: Evolution of the mass (left) and velocity (right) focused in the sand tank using different friction angle betwen Formica ($\delta_f$ = 23.5°) and urethane ($\delta_u$ = 19.8°)

# THE 1995 SHUM WAN LANDSLIDE, HONGKONG

**Description of the Simulation**

Analysis of Shum Wan Road landslide has been carried out using Coulomb-type friction law. The initial volume in our simulation is $26 \times 10^3 \, m^3$. Using griding algorithms (e.g. kriging), we get a 2 m grid spacing. The grid is 113 x 85 points. We performe several analysis dealing with Coulomb-type friction angle $\delta \in [16°, 26°]$.

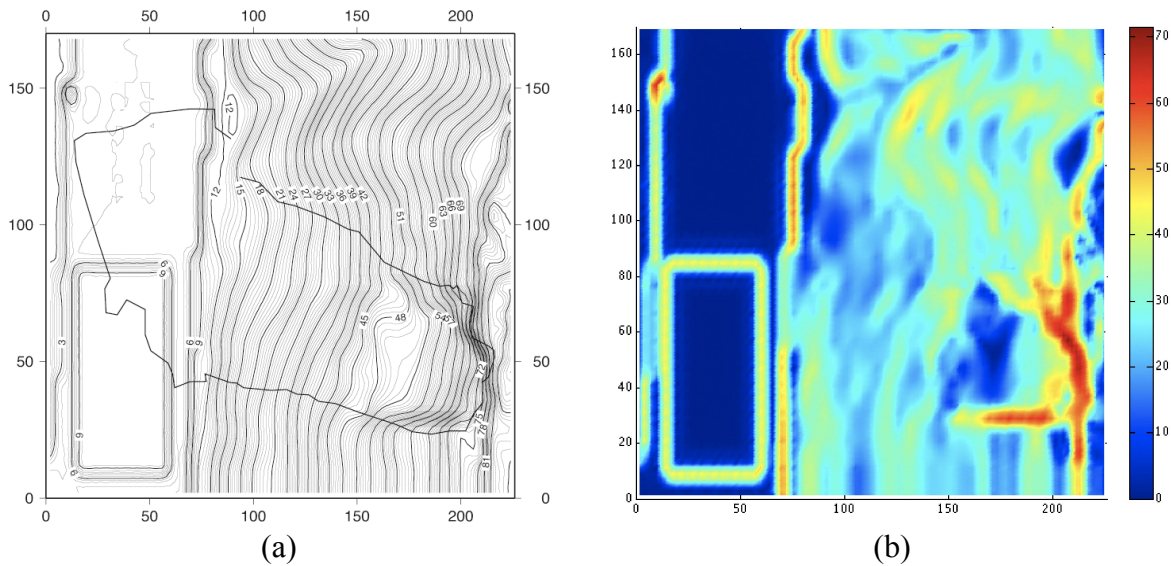

Figure 9: (a) DEM of Shum Wan landslide, HK. Slide path is indicated by black lines over the topographic map. (b) Slope map calculated from the DEM.

**Results**

The friction angle is calibrated to fit the runout distance of the landslide. Best agreement is obtained using $\delta = 18°$. The mean velocity increases with a factor of two during the first 10 seconds. The flow stops at $t = 30$ s.

Table 1: Velocity evolution and statistics for Shum Wan landslide

| Time (sec.) | $\bar{u}$ (m/sec) | $U_{max}$ (m/sec) | $\sigma$ | $\sigma/\bar{u}$ |
|---|---|---|---|---|
| 4 | 2.39 | 11.18 | 2.71 | 1.14 |
| 16 | 5.16 | 16.53 | 4.85 | 0.94 |
| 28 | 2.56 | 13.35 | 4.17 | 1.62 |

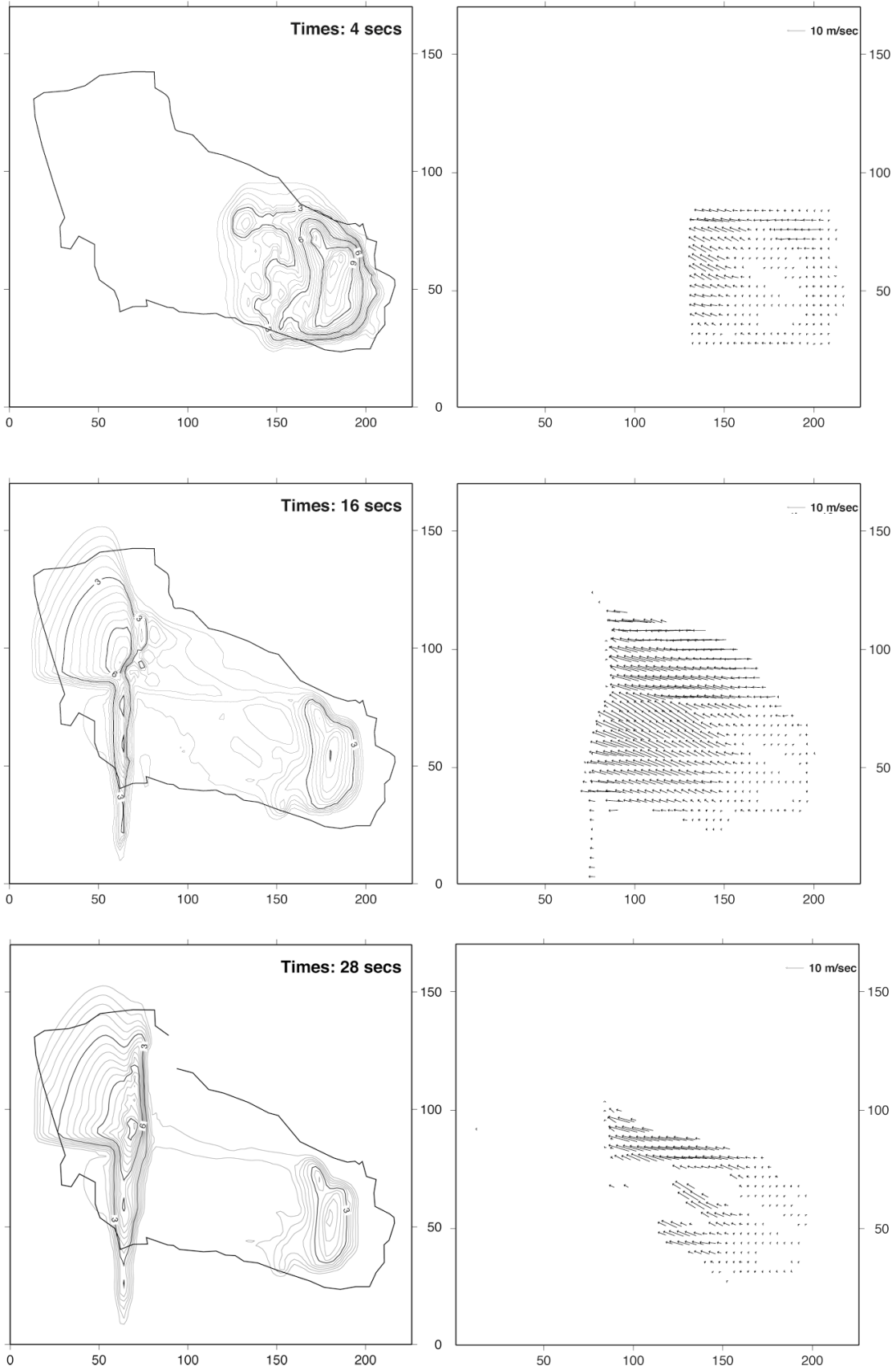

Figure 10: Evolution of the mass (left) and velocity field (right). Results from simulation using friction angle δ = 18°

# THE 1995 FEI TSUI LANDSLIDE, HONG KONG

**Description of the Simulation**

Analysis of Fei Tsui landslide has been carried out using Coulomb-type friction and Pouliquen friction laws. The initial volume in our simulation is $14 \times 10^3$ m$^3$ with a 1m grid spacing. The grid is 97 x 117 points. A series of simulation is performed using friction angles $\delta \in [22°, 30°]$.

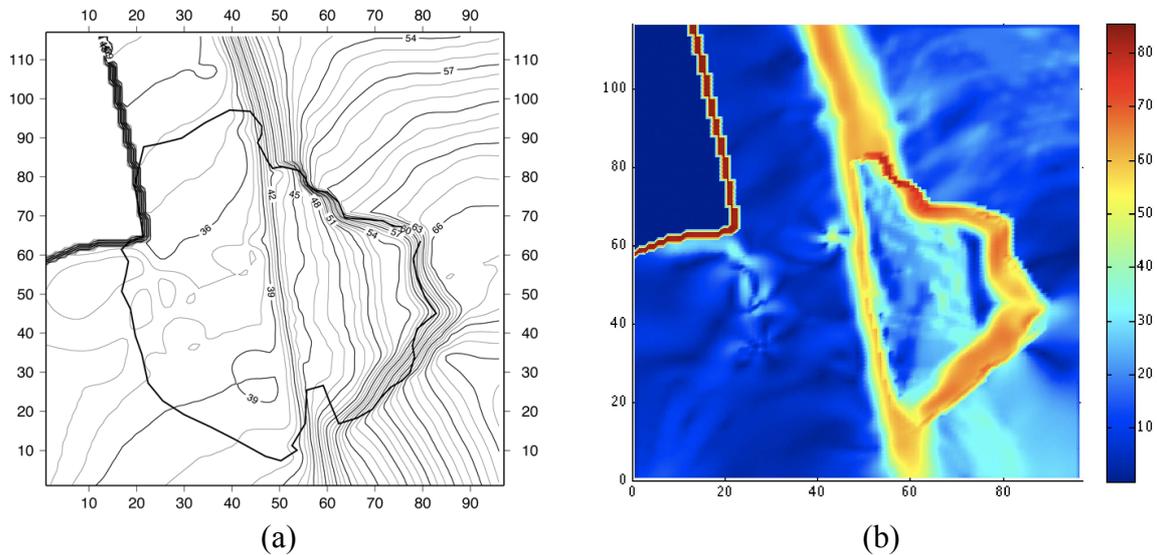

Figure 11: (a) DEM of Shum Wan landslide, HK. Slide path is indicated by black lines over the topographic map. (b) Slope map from the DEM.

**Results**

Better results are obtained with $\delta = 26°$ as it is shown in Figure 12. On the right side of Figure 12, we show the evolution of the velocity field at several times. The mean velocity increases rapidly after triggering and rise up to 70 m/s at its maximum. The flow stops at $t \approx$ 15 s (see Table 2).

Finest analysis shows that, between $t = 8$ s and $t = 14$ s, the back of the mass is still under movement whereas the front does not.

As consequences, the flow spread on each side (along the road) and not in the slope direction. We thus overestimate the width of the final deposits.

Table 2: Velocity evolution and statistics for Fei Tsui landslide

| Time (sec.) | $\bar{u}$ (m/sec) | $U_{max}$ (m/sec) | $\sigma$ | $\sigma/\bar{u}$ |
|---|---|---|---|---|
| 2 | 4.50 | 29.05 | 3.40 | 0.76 |
| 8 | 2.38 | 68.33 | 3.08 | 1.29 |
| 14 | 1.34 | 47.39 | 2.35 | 1.75 |

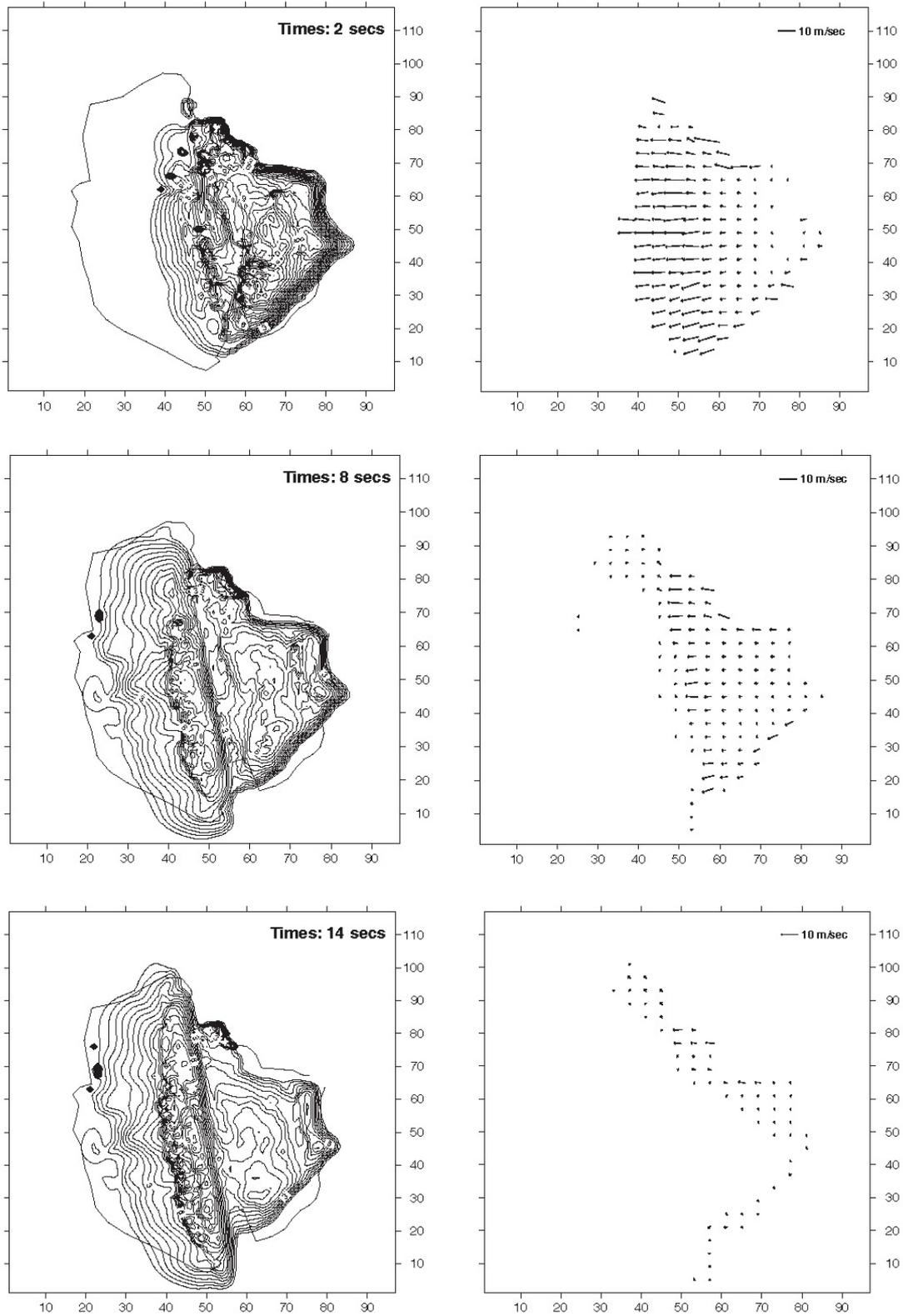

Figure 12: Evolution of the flow height and velocity at serveral times using Coulomb type friction angle δ = 26°. Isovalues of the thickness are represented every 1 m. On velocity held, only 1/4 vectors are displayed for a better visualization.

## THE 1902 FRANK SLIDE, CANADA

**Description of the Simulation**
The Frank Slide involves 36 x $10^6$ m$^3$. The grid is 201 x 201 points with 20 m space increment.

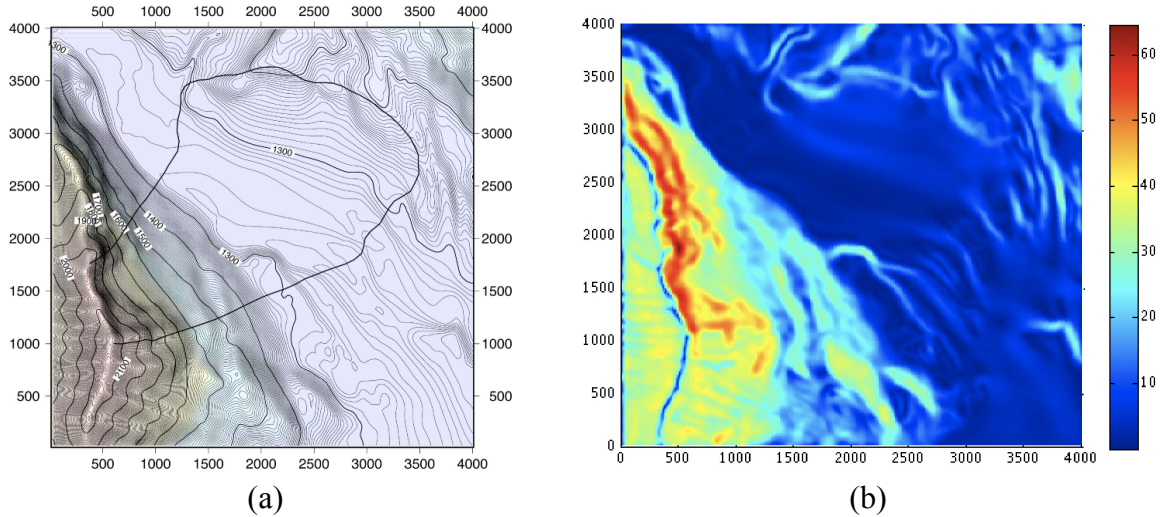

Figure 13: (a) DEM of Frank Slide, Canada. Slide path is indicated by black lines over the topographic map. (b) Slope map calculated from the DEM.

**Results**
Pirulli & Mangeney (2007) proposed a Coulomb-type friction angle $\delta = 14°$ for this specific landslide. Using the same angle *Shatop-2d* overestimates the observed runout as it is shown in Figure 14. We get better results with $\delta = 12°$ (Figure 15). Using a Pouliquen-type friction law, we get very similar results (Figure 16). Sensitivity of the numerical results to mesh size has to be performed before getting any conclusion from this comparison. Actually, the differences could only result from numerical dissipation (see discussion in the analytical solution section).

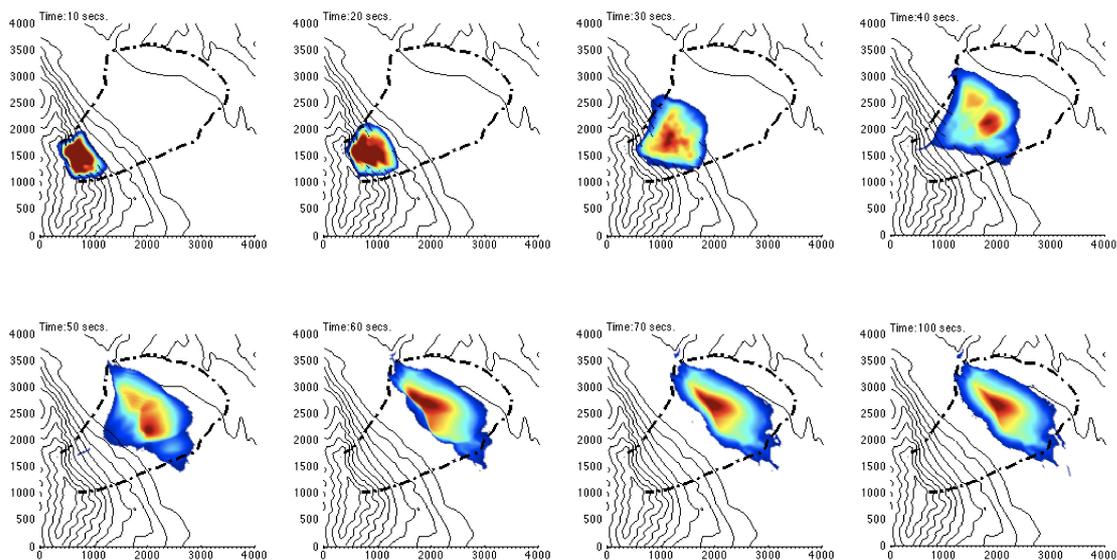

Figure 14: Evolution of the slide using friction angle $\delta = 14°$

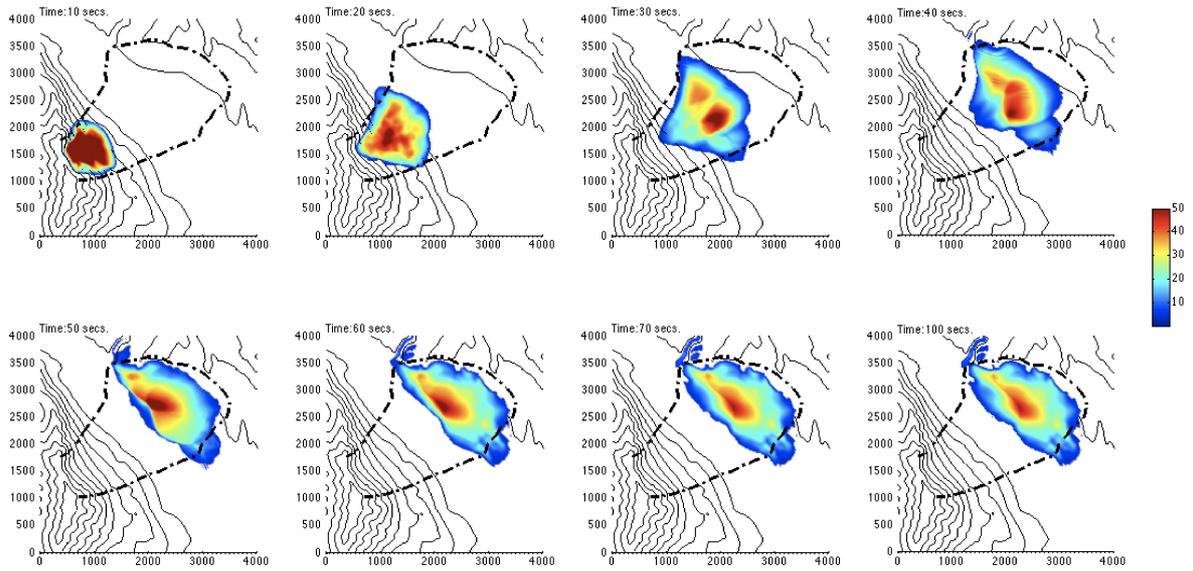

figure 15: Evolution of the slide using friction angle $\delta = 12°$

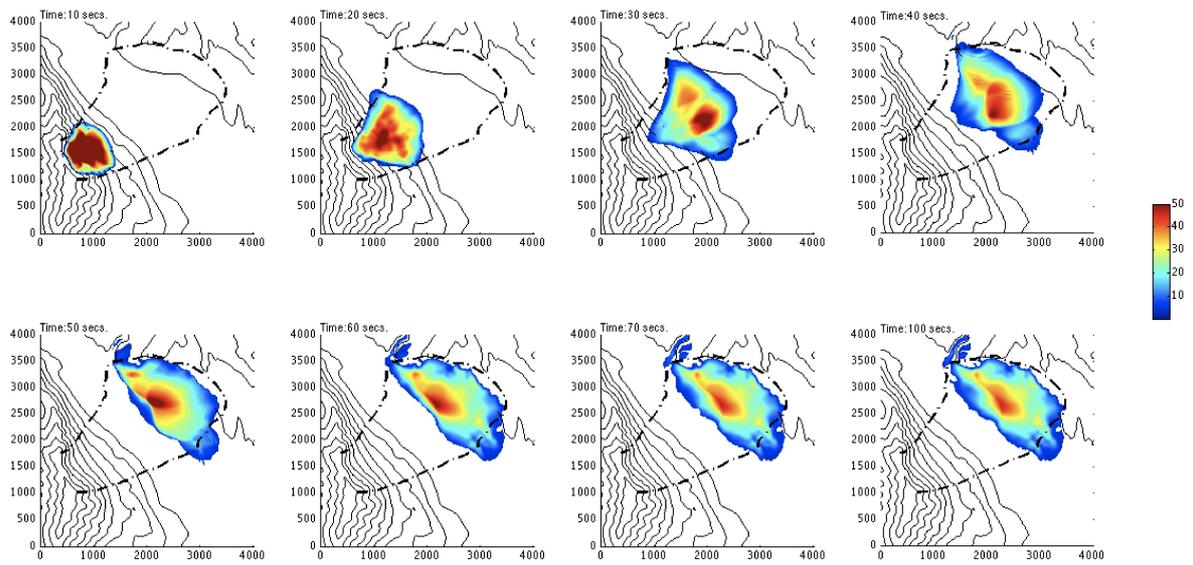

Figure 16: Evolution of the slide using a Pouliquen friction law where $\delta_1 = 12°$, $\delta_2 = 26°$ and $d = 1.5$m

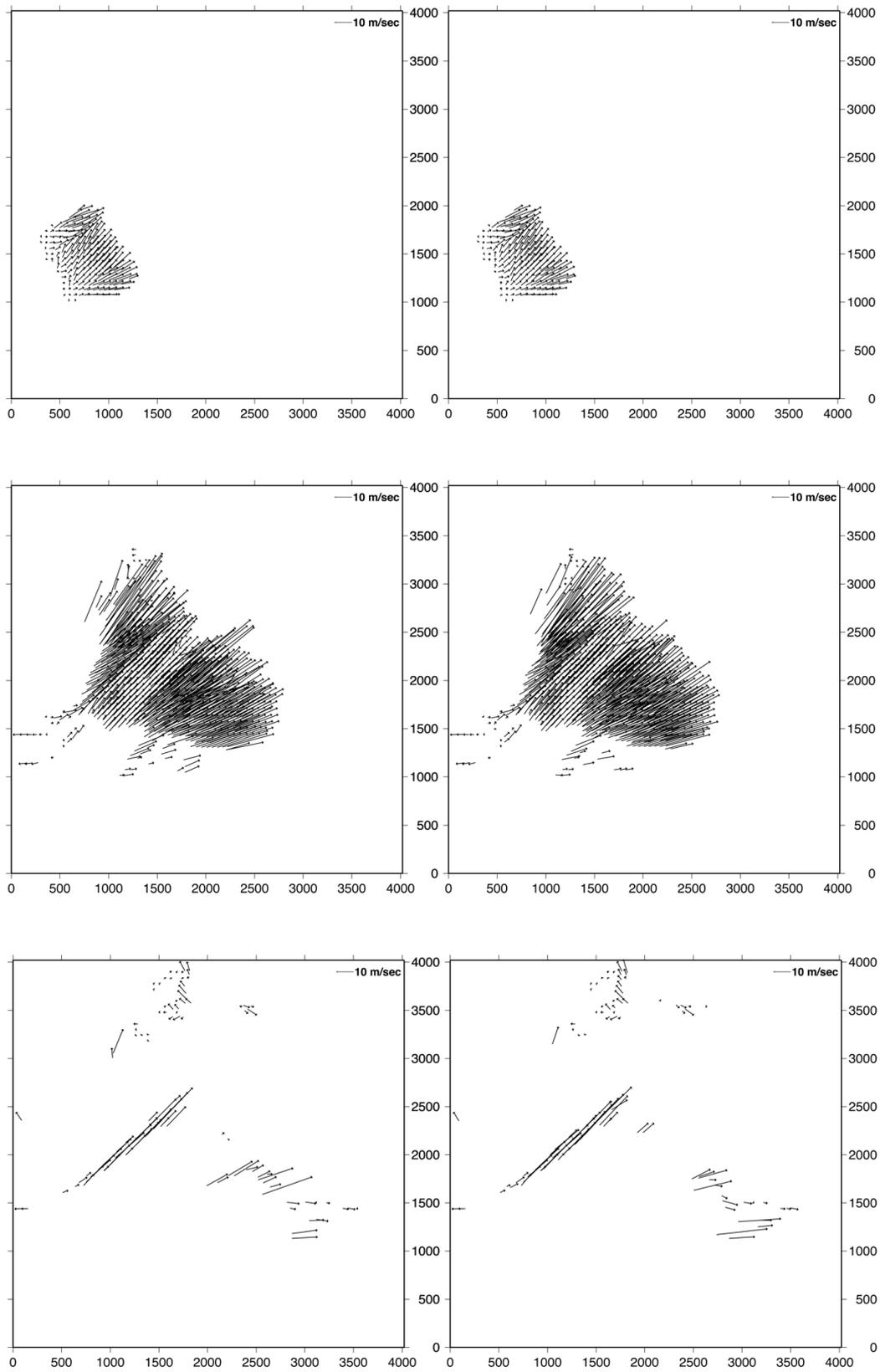

Figure 17: Velocity field at time = 10, 40 and 70 secs using Coulomb-type friction (left) and using Pouliquen-type friction law (right)

Using Coulomb-type friction or Pouliquen-type friction law have consequences on velocity field. With the second law, the flow gets faster but slows down quickly when compared with Coulomb-type friction law (see Table 3).

Table 3: Velocity evolution and statistics for both behaviors

| Coulomb-type Friction | | | | |
|---|---|---|---|---|
| Time (sec.) | $\bar{u}$ (m/sec) | $U_{max}$ (m/sec) | $\sigma$ | $\sigma/\bar{u}$ |
| 10 | 7.25 | 15.31 | 3.71 | 0.51 |
| 40 | 20.26 | 67.00 | 13.54 | 0.67 |
| 70 | 3.85 | 69.80 | 8.85 | 2.30 |
| Pouliquen-type Friction | | | | |
| Time (sec.) | $\bar{u}$ (m/sec) | $U_{max}$ (m/sec) | $\sigma$ | $\sigma/\bar{u}$ |
| 10 | 7.30 | 15.16 | 3.81 | 0.55 |
| 40 | 20.80 | 76.14 | 13.50 | 0.65 |
| 70 | 3.97 | 47.25 | 8.65 | 2.17 |

**CONCLUSIONS**

Our simulations show that, despite the accordingly non-physical empirical friction coefficient used in the numerical model, the area of the deposit is quite well reproduced in many natural cases. However, we showed that this coefficient ranges between 12° and 28° depending on the studied case. This dispersion of the friction coefficient questions the predictable power of this approach. The presence of a fluid phase, granulometry distribution, the presence of tree roots at the top of the released mass, as this is the case for Shum Wan landslide or Fei Tsui Road landslide, could explain the difference between the friction coeffcient. Only calibration in the similar geological context would possibly make it possible to predict further events.

Using both friction laws (Coulomb and Pouliquen) may have inferences on velocity field. In terms of prediction, this is a key point, because of the pressure impact on the foundations.

**REFERENCES**
Audusse, E., Bouchut, F., Bristeau, M. O., Klein, R., and Perthame, B. (2004). "A fast and stable wellbalanced scheme with hydrostatic reconstruction for shallow water flows." *SIAM J. Sci. Comp.*, 25, 2050-2065.
Bouchut, F. (2004). "Nonlinear stability of finite volume methods for hyperbolic conservation laws, and well-balanced schemes for sources." *Frontiers in Mathematics series, Birkhäuser.*
Bouchut, F., Mangeney-Castelnau, A., Perthame, B., and Vilotte, J. P. (2003). "A new model of Saint Venant and Savage-Hutter type for gravity driven shallow water flows." *C.R. Acad. Sci., Paris, série* I 336, 531-536.
Bouchut, F., and Westdickenberg, M. (2004). "Gravity driven shallow water models for arbitrary topography." *Comm. in Math. Sci.*, 2, 359-389.
Bristeau, M. O., Coussin, B. And Perthame, B. (2001). *Boundary Conditions for the Shallow Water Equations Solved by Kinetic Schemes*. INRIA Report, 4282.
Iverson, R. M., Logan, M., and Denlinger, R. P. (2004). "Granular avalanches across irregular threedimensional terrain: 2. Experimental tests." *J. Geophys. Res.*, 109,

F01015.

Lucas, A., and Mangeney, A. (2007). "Mobility and topographic effects for large Valles Marineris landslides on Mars." *Geophys. Res. Lett.*, 34, L10201, doi:10.1029/2007GL029835.

Mangeney-Castelnau, A., Vilotte, J. P., Bristeau, M. O., Perthame, B., Bouchut, F., Simeoni, C., and Yernini, S. (2003). "Numerical modeling of avalanches based on Saint-Venant equations using a kinetic scheme." *J. Geophys. Res.*, 108 (B11), 2527.

Mangeney-Castelnau, A., Bouchut, F., Vilotte, J. P., La jeunesse, E., Aubertin, A., and Pirulli, M. (2005). "On the use of Saint-Venant equations for simulating the spreading of a granular mass." *J. of Geophysical Research*, 110(B9), B09103.

Mangeney, A., Staron, L., Volfson, D., and Tsimring, L. (2006). "Comparison between discrete and continuum modeling of granular spreading." *WSEAS Transactions on Mathematics*, 2(6), 373-380.

Mangeney, A., Bouchut, F., Thomas, N., Vilotte, J. P., and Bristeau, M. O. (2007). "Numerical modeling of self-channeling granular flows and of their levee/channel deposits." *to appear in J. of Geophysical Research -Earth Surface.*

Pirulli, M. (2004). *Numerical Modelling of Landslide Runout, A Continuum Mechanics Approach.* PhD. Thesis in Geotechnical Engineering. Supervisors: Claudio Scavia, Politecnico di Torino; Anne Mangeney, IPGP-Université Paris-Diderot, France; Oldrich Hungr, University British Columbia, Vancouver.

Pirulli, M., Bristeau, M. O., Mangeney, A., and Scavia, C. (2007). "The effect of the earth pressure coefficients on the runout of granular material." *Environmental Modelling and Software*, 22, 1437-1454.

Pirulli, M., and Mangeney, A. (2007). "Result of back-analysis of the propagation of rock avalanches as a function of the assumed rheology." *Rock Mech. Rock Engng.*, DOI 10.1007/s00603-007-0143-x.

Pouliquen, O. (1999). "Scaling laws in granular flows down rough inclined planes." *Phys. Fluids*, 11, 542-548.

Wessel, P., and Smith, W. H. F. (1991). "Free software helps map and display data." *EOS Trans. AGU*, 72, 441.